\documentclass[10pt,prl,letterpaper,typeset,superscriptaddress,twocolumn,showpacs,spanish,english]{revtex4-1}
\usepackage{amsfonts}
\usepackage{amsmath}
\usepackage{amssymb}
\usepackage{graphicx}

\begin{document}


\title{Analogue of the quantum total probability rule from Paraconsistent  bayesian probability theory}

\author{R. Salazar}
\affiliation{Center for Optics and Photonics, Universidad de Concepci\'on, Casilla 4012, Concepci\'on, Chile}
\affiliation{MSI-Nucleus on Advanced Optics, Universidad de Concepci\'on, Casilla 160-C, Concepci\'on, Chile}
\author{C. Jara-Figueroa}
\affiliation{Center for Optics and Photonics, Universidad de Concepci\'on, Casilla 4012, Concepci\'on, Chile}
\affiliation{MSI-Nucleus on Advanced Optics, Universidad de Concepci\'on, Casilla 160-C, Concepci\'on, Chile}
\author{A. Delgado}\email{aldo.delgado@cefop.cl}
\affiliation{Center for Optics and Photonics, Universidad de Concepci\'on, Casilla 4012, Concepci\'on, Chile}
\affiliation{MSI-Nucleus on Advanced Optics, Universidad de Concepci\'on, Casilla 160-C, Concepci\'on, Chile}
\affiliation{Departamento de F\'isica, Universidad de Concepci\'on, Casilla 160-C, Concepci\'on, Chile}

\begin{abstract}
We derive an analogue of the quantum total probability rule by constructing a probability theory based on paraconsistent logic. 
Bayesian probability theory is constructed upon classical logic and a desiderata, that is, a set of desired properties that the 
theory must obey. We construct a new probability theory following the desiderata of Bayesian probability theory but replacing 
the classical logic by paraconsistent logic. This class of logic has been conceived to handle eventual inconsistencies or 
contradictions among logical propositions without leading to the trivialisation of the theory. Within this Paraconsistent bayesian probability theory it is possible to deduce a new total probability rule which depends  on the probabilities 
assigned to the inconsistencies. Certain assignments of values for these probabilities lead to expressions identical to those of 
Quantum mechanics, in particular to the quantum total probability rule obtained via symmetric informationally complete positive-operator valued measure.
\end{abstract}

\date{\today}
\pacs{03.65.Ta, 03.65.Wj, 03.67.-a}
\maketitle

According to the bayesian approach, probability theory is an extension of logic \cite{God}. Probabilities are a measure, assigned by an agent to the plausibility of a proposition conditional on the truth of a priori information \cite{Finetti}. In the limit of complete knowledge, when 
probabilities achieve extremal values, the rules for handling probabilities become those of deductive classical logic. 

Bayesian probability theory (BPT) has been successfully applied in a wide range of research areas such as for instance genetics \cite{genetics}
and cosmology \cite{cosmology}. In spite of this, there are phenomena which are beyond the scope of BPT. A noteworthy example is Quantum mechanics, 
where probabilities do not follow, in general, the total probability rule of BPT. Quantum states can be expressed as convex combinations of operators,
in a symmetrically informationally complete-positive operator valued measure (SIC-POVM), weighted by probabilities. This representation, together with Born's rule, allows us to calculate the probability of the outcomes associated to any other positive operator-valued measure (POVM), which 
leads to the quantum total probability rule. This turns out to be related to the total probability rule of BPT through rescaling and translation operations.

Arises thus naturally the question whether we can modify BPT in such a way that it provides us a new total probability rule which encompasses as 
particular cases the total probability rule of BPT and its quantum counterpart. Here, we attempt to answer this question by constructing a probability 
theory based on paraconsistent logic. Paraconsistent logics \cite{daCosta1,daCosta2,Priest} are designed to handle theories in which inconsistencies or contradictions might arise without leading to a trivialisation or logical explosion. Thereby, this new Paraconsistent bayesian probability theory (PBPT) requires the assignation of probabilities to the occurrence of contradictions. These enter in the new total probability rule and, depending on their values, it is possible to recover the total probability rule of BPT and its quantum version. Let us note that we do not seek a new interpretation of Quantum mechanics, but a new probability theory upon which we can build Quantum mechanics.

In BPT the basic product and sum rules for combining probabilities are given by
\begin{equation}
P(A,B|I)=P(A|I)P(B|A,I)=P(B|I)P(A|B,I)
\label{Product}
\end{equation}
and
\begin{equation}
P(A|I)+P(\bar A|I)=1,
\label{Sum}
\end{equation}
respectively. Symbols $A$, $B$ and $I$ represent propositions asserting that something is true and a bar over a proposition indicates 
its logical negation. A proposition to the right of a vertical bar is assumed to be true and a comma separating two propositions 
indicates the logical conjunction. Sum of two propositions indicates the logical disjunction. Thus $P(A|B,I)$ means the probability 
that $A$ is true conditional on the truth of both $B$ and $I$. Bayes rule follows by rearranging the two terms at the right hand 
side of the product rule, that is
\begin{equation}
P(B|A,I)=\frac{P(B|I)P(A|B,I)}{P(A|I)}.
\label{Bayes}
\end{equation}
The total probability rule, which follows from a marginalisation process over a set of complete, mutually exclusive propositions $\{A_i\}$, is given by
\begin{equation}
P(B|I)=\sum_i P(A_i|I)P(B|A_i,I).
\label{Tprobability1}
\end{equation} 
Sum and product rules can be deduced from classical logic by means of a desiderata \cite{God}, that is, a set of desirable properties that
a theory for plausible reasoning or inference should satisfy. In this regard, the desiderata does not assert anything to be true. 
The desiderata is: (i) degrees of plausibility are represented by real numbers, (ii) as new information supporting the truth of a 
proposition is supplied, the number which represents the plausibility will increase continuously and monotonically and achieve the 
deductive limit where appropriate, (iiia) if a conclusion can be reasoned out in
more than one way, every possible way must lead to the same result (structural consistency), (iiib) the theory must take account 
of all information, provided it is relevant to the question (propriety), and (iiic) equivalent states of knowledge must be represented 
by equivalent plausibility assignments (Jaynes consistency).

In order to construct the PBPT we choose to maintain the desiderata but change the underlaying logic. Instead of basing our construction 
on classical logic we resort to a paraconsistent logic, in particular the $\mathcal{C}_1$ propositional calculus \cite{daCosta3}. Paraconsistent logics are logics in 
which theories can be inconsistent but non-trivial. A trivial theory is one, such that everything expressed in its language can be proved. In classical logic,
any inconsistency entails triviality, that is: $(A,\bar A)\vdash B$ ( where $\vdash$ indicates syntactic consequence 
or formal deduction) for any formulas $A$ and $B$, which is not true in paraconsistent logics.

A theory {\it T}, whose underlying logic is {\it L} and whose language is $\mathcal{L}$, is {\it inconsistent}
if there is a formula $\alpha$ (an admisible expression of its language ) such that both $\alpha$ and $\bar\alpha$ are theorems (formulas deduced 
from the axioms of the theory by means of its rules of inference) of {\it T}, otherwise {\it T} is {\it consistent}. An expression of the form 
$\alpha,\bar\alpha$ is called {\it contradiction}. {\it T} is {\it trivial} if all formulas of $\mathcal{L}$ are theorems of {\it T}; otherwise, 
{\it T} is {\it non-trivial}. The logic {\it L} is {\it paraconsistent} if it can be the underlying logic of inconsistent but non-trivial theories.
The propositional calculus $\mathcal{C}_1$ contains the usual connectives: implication ($\alpha\to\beta$), conjunction ($\alpha,\beta$), 
disjunction ($\alpha+\beta$) and negation ($\bar\alpha$).

Two important properties of $\mathcal{C}_1$ are: (i) in general, the principle 
of non-contradiction does not hold, and (ii) from two contradictory propositions, that is one being the negation of the other, it is not 
possible to deduce any arbitrary third proposition. This latter property ensures that the presence of contradictions by no means entail the trivialisation 
of the theory. Within $\mathcal{C}_1$ we define the {\it non-contradictoriness} of a proposition $\alpha$ by $\alpha^o=\overline{\alpha,\bar\alpha}$.
If $\alpha^o$ is true then for $\alpha$ holds the principle of non-contradiction and we said that $\alpha$ is {\it non-contradictory}.
Otherwise we say that $\alpha$ is {\it contradictory}. The proposition $\overline{\alpha^o}$ is called the {\it contradictoriness} of 
$\alpha$. Other important definition is the {\it strong negation} of $\alpha$ by $\bar\alpha^*=\bar\alpha,\alpha^o$. Two important 
results follow from these definitions: (a) for any $\alpha$ we have that $\vdash(\alpha^o)^o$ (Arruda's theorem \cite{daCosta3}) and (b) for any $\alpha$ and $\beta$ 
the connectives $\alpha\to\beta$, $\alpha+\beta$, $\bar\alpha^*$ and $\alpha,\beta$ satisfy all schemas and inference rules of 
classical propositional calculus.

Le us now construct the PBPT. We will keep the desiderata used to construct BPT. However, for the desiderata (ii) the deductive
limit is that of the valid schemas of $\mathcal{C}_1$-logic, which is the same as the classical logic when the statements are 
non-contradictory. We must also consider that the contradictoriness of a statement is a new relevant information for the 
plausibility of the statement. In BPT the product rule is deduced \cite{Tribus} by analysing all possible functional forms 
for the probability. All but one of these are demonstrated to be inadequate by studying several particular cases such as
$A=B$, $I=A$, $I=\bar{A}$, etc.  These particular cases are included in $\mathcal{C}_1$-logic, provided that the contradictoriness
of the statement is a relevant information. Thus, by desiderata (iiib) a proposition like $I = \bar{A}$ should read as
$I = \bar{A},I$ or $I = \bar{A}, \overline{A^o}$ or $I = \bar{A}, A^o$ and the last one rules out the same functional 
form as $I = \bar{A}$. Since the deduction of the product rule Eq. (\ref{Product}) does not involve any particular schema
(only properties of the connectives, like the symmetry of conjunction) it aso holds for $\mathcal{C}_1$-logic.

For any non-contradictory formula in $\mathcal{C}_1$ all schemas from classical logic are also valid. Thus, the results 
of BPT are also valid for these. Consequently, we have
\begin{equation}
P(A|A^o,I)+P(\bar{A}|A^o,I)=1.
\label{BPT}
\end{equation}
Following (a) we have that any formula's non-contradictoriness is always non-contradictory. Thus
\begin{equation}
P(A^o|I)+P(\overline{A^o}|I)=1, 
\label{FromARRUDA}
\end{equation}
and according to (b) we have for the strong negation that
\begin{equation}
P(A|I')+P(\bar{A^*}|I')=1.
\label{FromSTRONGNEGATION}
\end{equation}

We can now deduce the sum rule of PBPT. Let us start with Eq. (\ref{BPT}) and multiply it by $P(A^o|I)$, that is, 
\begin{equation}
P(A^o|I)P(A|A^o,I)+P(A^o|I)P(\bar{A}|A^o,I)=P(A^o|I).
\end{equation}
Making use of the product rule Eq. (\ref{Product}) we obtain
\begin{equation}
P(A|I)P(A^o|A,I)+P(\bar{A}|I)P(A^o|\bar{A},I)=P(A^o|I).
\end{equation}
Using Eq. (\ref{FromARRUDA}) and the product rule the previous expression becomes
\begin{equation}
P(A|I)+P(\bar{A}|I)-P(\overline{A^o},A|I)-P(\overline{A^o},\bar{A}|I)=1-P(\overline{A^o}|I).
\end{equation}
Considering that by the definition of contradictoriness we have $\overline{A^o},A=\overline{A^o}$ and $\overline{A^o},\bar{A}=\overline{A^o}$ we finally obtain the sum rule of PBPT
\begin{equation}
P(A|I)+P(\bar{A}|I)-P(\overline{A^o}|I)=1.
\label{SumPBPT}
\end{equation}
Clearly, the PBPT sum rule, not only involves a proposition and its negation, but also its contradictoriness. In general $P(\overline{A^o})$ does not vanish and thus $P(A)\neq 1-P(\bar A)$. Furthermore, the probability of a contradiction works as a negative probability when thinking about it as an statement itself. Consequently, the addition of $P(A|I)$ and $P(\bar{A}|I)$ might be larger than one. In particular, when we known with certainty that $A$ is contradictory, that is, $P(\overline{A^o})=1$, then the sum rule demands the assignment $P(A)=1=P(\bar{A})$.

From Eq. (\ref{FromSTRONGNEGATION}) and using Eq. (\ref{FromARRUDA}) together with product and sum rules Eqs. (\ref{Product}) and
(\ref{SumPBPT}) we can deduce the extended sum rule which involves the probability of the disjunction of propositions $A$ and $B$, that is,
\begin{equation}
P(A+B|I)=P(A|I)+P(B|I)-P(A,B|I).
\label{ExtendedSum}
\end{equation} 
This is the same rule as in the case of BPT.

When analysing the probability of an statement, such as $A$, we can also reason based on the knowledge about their non-contradictory
or contradictory parts, that is, on $\tilde A=A, A^o$ or $A, \overline{A^o}$, respectively. The propriety desiderata (iiib) demands that this information must
play a relevant role in the assignment of probability. Since $A,\overline{A^o} = \overline{A^o}$, a reasoning based on the contradictory
part never reaches the deductive classical limit (i.e. $I = A^o,I$), thus desiderata (iiib) suggests us to consider the non-contradictoriness 
of our data. For instance, we can deduce the probability $P(\tilde A+\tilde B|I)$ for the conjunction of the mutually exclusive non-contradictory parts of $A$ and $B$, that is when the proposition $A,A^o,B,B^o$ does not hold. Under this condition the extended sum rule Eq. (\ref{ExtendedSum}) leads to
\begin{equation}
P(\tilde A+\tilde B|I)=P(\tilde A|I)+P(\tilde B|I),
\label{SumConsistentParts}
\end{equation}
which with the product rule becomes
\begin{eqnarray}
P(\tilde A+\tilde B|I)&=&P(A|I)P(A^o|A,I)
\nonumber\\
&+&P(B|I)P(B^o|B,I).
\end{eqnarray}
Eq. (\ref{FromARRUDA}) allows us to transform the terms involving the non-contradictoriness in the previous expression into terms involving the contradictoriness, that is
\begin{eqnarray}
P(\tilde A+\tilde B|I)&=&P(A|I)[1-P(\overline{A^o}|A,I)]
\nonumber\\
&+&P(B|I)[1-P(\overline{B^o}|B,I)].
\end{eqnarray}
Employing once again the product rule and  $A,\overline{A^o} = \overline{A^o}$ and $B,\overline{B^o} = \overline{B^o}$ we finally obtain the expression
\begin{eqnarray}
P(\tilde A+\tilde B|I)&=&P(A|I)+P(B|I)
\nonumber\\
&-&[P(\overline{A^o}|I)+P(\overline{B^o}|I)].
\label{ExtendedSumNon-contradictory}
\end{eqnarray}
Thus, the extended sum rule for two mutually exclusive non-contradictory parts turns out to be fundamentally different to Eq. (\ref{ExtendedSum}). The last term in Eq. (\ref{ExtendedSumNon-contradictory}) corresponds to the negative of the total probability that the statements are contradictory. Equivalently, when 
considering the sum rule of PBPT Eq. (\ref{ExtendedSumNon-contradictory}) becomes
\begin{equation}
P(\tilde A+\tilde B|I)=2-P(\bar{A}|I)-P(\bar{B}|I).
\end{equation}
Eq. (\ref{ExtendedSumNon-contradictory}) can be easily extended to the conjunction of $N$ propositions $A_k$ (with $k=1,\dots,N$) with mutually exclusive non-contradictory parts, that is, 
\begin{equation}
P(\sum_{k}\tilde A_k|I)=\sum_{k} P(A_k|I)-\sum_{k} P(\overline{A_k^o}|I).
\label{ExtendedSumNon-contradictory-N}
\end{equation}
This later result allows us to calculate the probability of the conjunction between propositions $B$ and $\sum_k\tilde A_k$. The product rules leads us to
\begin{equation}
P(B|\sum_k\tilde A_k,I)P(\sum_k\tilde A_k|I)=P(B,\sum_k\tilde A_k|I),
\end{equation}
which when combined with Eq. (\ref{ExtendedSumNon-contradictory-N}) leads to
\begin{equation}
P(B|\sum_k\tilde A_kI)=\frac{P(B,\sum_k\tilde A_k|I)}{\sum_{k} [P(A_k|I)-P(\overline{A_k^o}|I)]}.
\end{equation}
Using Eq. (\ref{SumConsistentParts}) we obtain
\begin{equation}
P(B|\sum_k\tilde A_k,I)=\frac{\sum_kP(\tilde A_k|I)P(B|\tilde A_k,I)}{\sum_{k} [P(A_k|I)-P(\overline{A_k^o}|I)]}.
\label{B}
\end{equation}
Decomposing a proposition $A_k$ into its contradictory and non-contradictory parts we have $A_k=\tilde A_k+A_k,\overline{ A^o_k}$, or equivalently $A_k=\tilde A_k+\overline{A^o_k}$. Using the extended sum rule Eq. (\ref{ExtendedSum}) we obtain
\begin{equation}
P(A_k|I)=P(\tilde A_k|I)+P(\overline{A^o_k}|I)-P(\tilde A_k,\overline{A^o_k}|I).
\end{equation}
The last term vanishes since proposition $A^o_k$ is non-contradictory. Thereby, we have
\begin{equation}
P(A_k|I)=P(\tilde A_k|I)+P(\overline{A^o_k}|I).
\label{EQ1}
\end{equation}
This result allows us to cast Eq. (\ref{B}) as
\begin{eqnarray}
P(B|\sum_k\tilde A_k,I)&=&\frac{\sum_kP(A_k|I)P(B|\tilde A_k,I)}{\sum_{k} [P(A_k|I)-P(\overline{A_k^o}|I)]}
\nonumber\\
&-&\frac{\sum_kP(\overline{A_k^o}|I)P(B|\tilde A_k,I)}{\sum_{k} [P(A_k|I)-P(\overline{A_k^o}|I)]},
\label{B1}
\end{eqnarray}
which is the total probability rule of PBPT. Let us now consider the propositions ${A_k}$ to be non-contradictory, that is, our a priori information is $I=A_1^o,A_2^o,\dots,A_N^o,I$. Since propositions $\{\tilde A_k\}$ are mutually exclusive, we also have that propositions $\{A_k\}$ are mutually exclusive. Assuming the completeness of the set $\{A_k\}$ we have that $P(\sum_kA_k|I)=1=\sum_k P(A_k|I)$. Thereby, the second term at the right hand side of Eq. (\ref{B1}) vanishes and the denominator becomes 1. We obtain
\begin{equation}
P(B|\sum_k A_k,I)=\sum_k P(A_k|I)P(B|A_k,I).
\label{TProbabilityPBPT}
\end{equation}
Thus, the total probability rule of BPT given by Eq. (\ref{Tprobability1}) is contained within PBPT.  Total probability rule of PBPT Eq. (\ref{B1}) also allows to calculate probabilities $P(\overline B|\sum_k\tilde A_k,I)$ and $P(\overline{B^o}|\sum_k\tilde A_k,I)$, which together with probability $P(B|\sum_k\tilde A_k,I)$ also obey the sume rule Eq. (\ref{SumPBPT}).

In order to obtain the total probability rule of BPT we assumed the non-contradictoriness of propositions $\{A_k\}$. We can now show that the total probability rule Eq. (\ref{B1}) contains the quantum total probability rule for a set of propositions endowed with a particular structure of contradictoriness. Let us first briefly review the quantum total probability rule. Quantum states of a $d$-dimensional quantum system are described by unit-trace positive semidefinite linear operators $\rho$ which act onto the Hilbert space ${\cal H}_d$. States can also be described by a collection of probabilities with the help of a SIC-POVM \cite{Renes}. This is composed of $d^2$ subnormalised rank one projectors $\{\Pi_k/d\}$ with Hilbert-Schmidt products given by $Tr(\Pi_k\Pi_l)=(d\delta_{k,l}+1)/(d+1)$. These operators generate the representation for quantum states
\begin{equation}
\rho=\frac{d+1}{d}\sum_{k=0}^{d^2-1}Q(\Pi_k|I)\Pi_k-\frac{1}{d}\sum_{k=0}^{d^2-1}\Pi_k,
\label{qstate}
\end{equation}
where $Q(\Pi_k|I)=Tr(\Pi_k\rho)$ is Born's rule \cite{Born}. This set of probabilities contains all the information about the system required to predict the outcomes of possible experiments. Using this representation for quantum states we can calculate the transition probability associated to any other state $\Sigma$, that is
\begin{eqnarray}
Q(\Sigma|\rho)&=&\frac{d+1}{d}\sum_{k=0}^{d^2-1}Q(\Pi_k|I)Q(\Sigma|\Pi_k)
\nonumber\\
&-&\frac{1}{d}\sum_kQ(\Sigma|\Pi_k).
\label{quantum-total-probability-rule}
\end{eqnarray}
This expression is the quantum total probability rule and allows us to predict the transition probability to another state $\Sigma$ from our knowledge of the initial state of the system, given by the sets of probabilities $\{Q(\Pi_k|I)\}$, and the probabilities $\{Q(\Sigma|\Pi_k)\}$. In this regard, the quantum total probability rule is equivalent to Born's rule \cite{Fuchs}. 

Quantum and paraconsistent total probability rules contain the subtraction of two non-negative terms being one of them a constant. With the introduction of a particular set of propositions, which emulates certain properties of a SIC-POVM,  we can deduce a paraconsistent total probability rule that exhibits a stronger similarity with the quantum total probability rule. From Eq. (\ref{EQ1}) we have
\begin{equation}
\sum_iP(A_i|I)=\sum_iP(\tilde A_i|I)+\sum_iP(\overline{A^o_i}|I).
\label{EQ2}
\end{equation}
We also have that
\begin{equation}
P(\sum_iA_i|I)=1=P(\sum_i\tilde A_i+\sum_i\overline{A_i^o}|I).
\end{equation}
We now assume that propositions $A_k$ are all simultaneously contradictory or non-contradictory. In this case we require a single proposition to signal this property, that is $\overline{A^o_1}=\overline{A^o_2}=\dots=\overline{A^o_N}=\overline{A^o}$. Thereby, we have that $P(\sum_i\tilde A_i+\overline{A^o}|I)=1$, which with the help of the extended sum rule Eq. (\ref{ExtendedSum}) leads to
\begin{equation}
\sum_iP(\tilde A_i|I)+P(\overline{A^o}|I)=1.
\label{EQ3}
\end{equation}
Inserting Eq. (\ref{EQ3}) into Eq. (\ref{EQ2}) we obtain 
\begin{equation}
P(\overline{A^o}|I)=\frac{1}{N-1}\sum_iP(A_i|I)-\frac{1}{N-1}.
\end{equation}
This latter equation allows us to write Eq. (\ref{B1}) in the form
\begin{eqnarray}
P(B|\sum_k\tilde A_k,I)&=&\frac{N-1}{N-S(I)}\sum_kP(A_k|I)P(B|\tilde A_k,I)
\nonumber\\
&-&\frac{S(I)-1}{N-S(I)}\sum_kP(B|\tilde A_k,I),
\label{PBPTTOYMODEL}
\end{eqnarray}
with $S(I)=\sum_kP(A_k|I)$. We can now link both total probability rules Eqs. (\ref{PBPTTOYMODEL}) and (\ref{quantum-total-probability-rule}). In our model of propositions we now make the number of propositions equal to the number of elements in the SIC-POVM, that is $N=d^2$, and assign the value $\sum_iP(A_i|I)=d$. Entering these values in Eq. (\ref{PBPTTOYMODEL}) we obtain the expression
\begin{eqnarray}
P(B|\sum_{k=0}^{d^2-1}\tilde A_k,I)&=&\frac{d+1}{d}\sum_{k=0}^{d^2-1}P(A_k|I)P(B|\tilde A_k,I)
\nonumber\\
&-&\frac{1}{d}\sum_{k=0}^{d^2-1}P(B|\tilde A_k,I).
\label{TProbability1PBPT}
\end{eqnarray}
The similarity between total probability rules Eqs. (\ref{quantum-total-probability-rule}) and (\ref{TProbability1PBPT}) is striking. In Quantum mechanics the set $\{Q(\Pi_k|I)\}$ provides all the information needed to associate the system to a unique quantum state $\rho$ through Eq. (\ref{qstate}). In PBPT the set $\{P(A_k|I)\}$ represents our state of knowledge about the truth values of the set $\{A_k\}$ of complete, possibly contradictory, and mutually exclusive propositions. In Quantum mechanics the quantities $\{Q(\Sigma|\Pi_k)\}$ represent the probabilities of projecting onto state $\Sigma$ provided that we know with certainty that the system is described by the state $\Pi_k$. This set characterises the state $\Sigma$ with respect to the members of the SIC-POVM. In PBPT the set $\{P(B|\tilde A_k,I)\}$ has an analogous meaning, it describes the probabilities for the truth of proposition $B$ given that each proposition $A_k$ is considered to be true and non-contradictory. These probabilities summarise our knowledge about proposition $B$ given our knowledge about propositions $\{A_k\}$. If we now chose to assign the values of $Q(\Pi_k|I)$ and $Q(\Sigma|\Pi_k)$ to the values of $P(A_k|I)$ and $P(B|\tilde A_k,I)$ respectively, then both rules Eqs. (\ref{quantum-total-probability-rule}) and (\ref{TProbability1PBPT}) lead to the same prediction for the value of the probabilities $Q(\Sigma|\rho)$ and $P(B|\sum_{k=0}^{d^2-1}\tilde A_k,I)$.  We have thus now two theories, Quantum mechanics and Paraconsistent bayesian probability theory, with differences in their origins and formalisms. Yet, with the proper identifications, these lead to the same predictions. 

In recent years there has been an increasing interest in determining whether other theories might exhibit properties considered to be hallmarks of Quantum mechanics such as for instance interference, entanglement and incompatible measurements; this in the hope that Quantum mechanics can be deduced from these more general theories with the help of a suitable set of physically motivated constraints. For example, Spekken's toy model \cite{Spekkens,vanEnk} was constructed upon the knowledge balance principle. Although the model has many properties in common with Quantum mechanics, this does not emerge as a particular case of it. Generalised probabilistic theories \cite{Barrett} are based on a set of seven assumptions which allow to construct states and operations for composite systems which include those of Quantum mechanics. Here, we have introduced a new theory of probability based on the sole premise that its underlying logic must account for the possibility of contradictions. Thus, PBPT is a quantitative formulation of how to make rational decisions in presence of uncertainties and contradictions. This new theory is clearly of epistemological character, it says more about our believes on the plausibility of nature's behaviour than about nature itself, that is, PBPT does not describe the physical reality, it rather provides a set of rules, or algorithm, to calculate probabilities of certain propositions. Whether these propositions are of ontological or epistemological character cannot be decided within PBPT.
In absence of contradictions PBPT reduces to BPT. Within PBPT and at hand of a simple model of propositions we deduced a total probability rule which agrees in form and meaning with the quantum mechanical one. This is remarkable, since we have only demanded a change of the underlying logic without resorting to any conception about physical reality. Nevertheless, this principle is not enough to single out a space state which agrees with the quantum mechanical one. In the case of Quantum mechanics operators representing states must be positive semidefinite, which imposes $d-1$ constraints in the set of probabilities $Q(\Pi_k|I)$. These constraints do not arise within PBPT and thus additional principles are required. 

Within the model of propositions leading to Eq. (\ref{TProbability1PBPT}) we obtain that the probability of a contradiction is given by $P(\overline{A^o}|I)=1/(d+1)$. This value equals the inner product among different members  of the SIC-POVM, that is, $Tr(\Pi_k\Pi_l)=1/(d+1)$. The relationship between nonorthogonality of quantum states and emergence of logical contradictions was noted by Birkhoff and von Neumann \cite{Birkhoff}. In order to solve the difficulties posed by these contradictions they proposed to change the rules of classical logic by modifying the distributive identities of disjunction and conjunction. In our approach we do not attempt to eliminate logical contradictions but incorporate them into the formalism by means of paraconsistent logic, in particular the propositional calculus $\mathcal{C}_1$. This choice is motivated by simplicity. Within this calculus all schemas and inference rules of classical propositional calculus are valid when replacing the negation by the strong negation and the consistency of the contradictoriness of any proposition always holds. These elements also appear in other propositional calculi, such as for example Paraconsistent Boolean algebra, and thus our results might hold for other paraconsistent logics. 

Finally, we would like to mention that the connection between Quantum mechanics and Paraconsistent logic has been mentioned in the literature. In particular, this has been studied in the context of the superposition principle \cite{daCosta4} and quantum computing \cite{Agudelo}. Application of PBPT to the problem of inconsistent data basis is also feasible.

\section{Acknowledgments}

Authors acknowledge funding from Grants PIA-CONICYT PFB08024, ICM P10-030F and FONDECYT $\rm{N}^o$ 1140635.

\end{document}